\documentclass{article}

\usepackage{arxiv}

\usepackage[utf8]{inputenc} 
\usepackage[T1]{fontenc}    
\usepackage{hyperref}       
\usepackage{url}            
\usepackage{booktabs}       
\usepackage{amsfonts}       
\usepackage{nicefrac}       
\usepackage{microtype}      
\usepackage[none]{hyphenat}
\usepackage{lipsum}
\usepackage{amsmath}
\usepackage{multirow}
\usepackage{stfloats}
\usepackage{graphicx}
\graphicspath{ {./images/} }

\title{Enhanced SegNet with Integrated Grad-CAM for Interpretable Retinal Layer Segmentation in OCT Images}

\author{
 S M Asiful Islam Saky \\
  School of Coumputing and Informatics\\
  Albukhary Internationation University\\
  Alor Setar, Kedah, Malaysia \\
  \texttt{asiful.saky@student.aiu.edu.my} \\
   \And
 Ugyen Tshering \\
  School of Coumputing and Informatics\\
  Albukhary International University\\
  Alor Setar, Kedah, Malaysia \\
  \texttt{ugyen.tshering@student.aiu.edu.my} \\
}

\begin{document}
\maketitle
\begin{abstract}
For the diagnosis and treatment of vision-threatening conditions such as glaucoma, diabetic retinopathy, and age-related macular degeneration, precise retinal layer segmentation in Optical Coherence Tomography (OCT) is essential. Despite their high technical accuracy, automated deep learning techniques frequently lack the interpretability needed for clinical trust. Meanwhile, manual segmentation remains time- consuming and inconsistent. The "black-box" nature of conventional models, pathological distortions, and image noise are major obstacles. An improved SegNet-based deep learning framework for automated, interpretable retinal layer segmentation is proposed in this work. Architectural innovations such as modified pooling strategies were integrated to enhance feature extraction from noisy OCT images. To address class imbalance and improve thin layer segmentation, a hybrid loss function that combines categorical Cross-Entropy and Dice Loss was developed. Crucially, Gradient-weighted Class Activation Mapping (Grad-CAM) was integrated into the pipeline to provide visual explanations of the model's decision-making process, enabling clinical validation. The model was trained and validated on the publicly available Duke OCT dataset. The proposed framework demonstrated high performance, achieving a validation accuracy of 95.77\%, Dice Coefficient of 0.9446, and Jaccard Index (IoU) of 0.8951, indicating robust and accurate segmentation across diverse scans. Class-wise analysis confirmed strong performance for most layers, though challenges remained for thinner layers with complex boundaries. Grad-CAM visualizations successfully highlighted anatomically relevant regions influencing segmentation decisions, confirming the model's alignment with clinical biomarkers and enhancing transparency. Using an Explainable AI (XAI) technique (Grad- CAM), a customized loss function, and architectural improvements, this study offers a strong SegNet-based framework that significantly enhances automated retinal layer segmentation. The model bridges the gap between algorithmic performance and clinical utility by achieving high accuracy while maintaining crucial interpretability. This approach holds significant potential for standardizing OCT analysis, improving diagnostic efficiency, and fostering trust in AI-driven ophthalmic tools.
\end{abstract}


\section{Introduction}
By allowing high-resolution, cross-sectional imaging of retinal layers, Optical Coherence Tomography (OCT) has transformed ophthalmology; this advancement is vital for diagnosing and controlling vision-threatening disorders including glaucoma, diabetic retinopathy (DR), and age-related macular degeneration (AMD). Early and accurate detection is essential to avoid irreversible vision loss since these diseases together impact more than 300 million individuals globally \cite{wong2016diabetic}. Accurate retinal layer segmentation in OCT produces quantitative biomarkers vital for clinical decisions; for example, retinal nerve fibre layer (RNFL) thickness is key in monitoring glaucoma \cite{leung2011evaluation}, while delineating features in DR and AMD guides intervention and treatment monitoring. Though clinically significant, manual segmentation of retinal layers is still time-consuming and variable. Research has indicated that manual RNFL readings could differ by more than 15\% between clinicians, potentially leading to different diagnostic conclusions \cite{mwanza2015residual}. Moreover, the procedure is labour-intensive and sometimes requires up to 30 minutes every scan, which is impractical for busy clinical environments \cite{kushner2020retinal}. With automated systems reaching comparable or better accuracy to manual techniques in controlled studies, recent developments in deep learning have shown the promise of artificial intelligence (AI) to solve these issues \cite{defauw2018clinically}. However, significant obstacles remain to their extensive clinical use.

Automated OCT segmentation has various technical and practical issues that limit its clinical usefulness. First, speckle noise—an inherent artefact of the OCT imaging process that obscures fine anatomical details and mimics diseased features—often degrades OCT images \cite{adhi2013optical}. Particularly in areas where contrast between adjacent layers is naturally low, such as the interface between the inner nuclear and outer plexiform layers, this noise makes it more difficult to identify layer borders. Second, pathological changes in conditions like AMD and DR add further intricacy. Subretinal fluid, drusen, or cystoid macular oedema, for instance, can distort retinal anatomy and cause segmentation errors in as many as 30\% of instances \cite{tang2021multitask}. Differences in OCT equipment and imaging techniques present another important challenge by way of unpredictability. Different manufacturers' commercial OCT systems use proprietary hardware and software, which causes differences in image resolution, signal-to-noise ratio, and scanning patterns. Research has indicated that segmentation models trained on data from one device frequently suffer performance declines of 15--25\% when applied to photos from another system, limiting their generalizability \cite{holzinger2019causability}. Perhaps, the most pressing challenge is the ``black-box'' nature of many deep learning models. Although convolutional neural networks (CNNs) such as U-Net have reached state-of-the-art performance in segmentation activities, their decision-making mechanisms still remain opaque to clinicians. Ophthalmologists cannot confirm whether segmentation findings match anatomical or pathological ground truths, so this lack of interpretability erodes confidence in automated systems.

Current OCT segmentation methods can be generally divided into two groups: deep learning-based approaches and conventional image-processing techniques. Among the first to automate retinal layer segmentation were traditional methods such as active contour models and graph-based algorithms—e.g., the Iowa Reference Algorithm \cite{garvin2009automated}. Although these techniques reduced dependence on manual tracing, they often required expert initialization and underperformed in pathological settings. Though computationally expensive and noise sensitive, level-set techniques enhanced border identification, making them less practical for routine clinical use. Deep learning has changed the industry; CNNs such as U-Net have attained Dice scores of 0.85--0.90 on standardised datasets \cite{ronneberger2015unet}. These models, meanwhile, have certain drawbacks. First, they need significant annotated datasets for training, which are expensive and labour-intensive to create \cite{gulshan2016development}. Furthermore, their performance often suffers markedly when applied to images from new devices or populations not included in the training data. Most importantly, these models typically lack inherent interpretability. A model segmented a layer in a certain manner; hence clinicians cannot understand its results  \cite{hassija2024interpreting}. Though few studies have effectively included interpretability straight into the segmentation process, recent work has tried to fill these gaps with transformer-based models and attention mechanisms. This is a crucial unmet requirement in the discipline since, ultimately, automated segmentation aims for clinical usefulness and acceptance rather than only technical correctness \cite{wang2021automated}.

To address these limitations, we proposed an improved SegNet-based deep learning framework that significantly improves automatic retinal layer segmentation in OCT images. Our main contributions include technical as well as clinical challenges in the sector. First, we have introduced architectural changes to the conventional SegNet model, including hybrid pooling indices and adjustable receptive fields. While preserving computational efficiency, these changes improve the model's capacity to capture tiny details in noisy OCT images. We have integrated a hybrid loss function combining categorical cross-entropy and Dice loss to solve the problem of class imbalance in retinal layers. Particularly for thinner structures that are more difficult to identify, this method increases segmentation accuracy across all retinal layers. Another significant advancement is the direct integration of Gradient-weighted Class Activation Mapping (Grad-CAM) into the segmentation pipeline. This explainability tool generates intuitive heat maps highlighting the anatomical areas affecting the model's decisions, allowing clinicians to validate the model's findings against their own anatomical knowledge and expertise. The proposed framework shows strong performance through comprehensive evaluation, achieving high scores on standard segmentation metrics. By combining improved accuracy with interpretability, this study bridges an important gap between research and clinical practice. The suggested paradigm has considerable promise to transform OCT analysis into a more efficient, standardised part of precision ophthalmology, hence improving patient care all around.

\section{Literature Review}
Optical Coherence Tomography (OCT) has become an essential diagnostic tool in ophthalmology that provides high-resolution cross-sectional images of the retina. This technique allows clinicians to study the complex structure of the retina in detail, which is key for the diagnosis and monitoring of conditions such as age-related macular degeneration (AMD), diabetic macular edema (DME), and glaucoma \cite{salz2016select}. Retinal layer segmentation has many applications, involving the identification of boundaries between different retinal layers, which allows for accurate measurement of their thickness and structure. Thus, segmentation plays an important role in early disease detection, monitoring disease progression, and treatment evaluation of medical imaging.

\subsection{Traditional Methods for Retinal Layer Segmentation in OCT}
Early methods for retinal layer segmentation primarily relied on manual or semi-automated approaches. While usually accurate, these methods were laborious and impractical for routine use in busy clinical settings. Traditional methods included graph-based algorithms and classical machine learning approaches, which often relied heavily on preprocessing steps and manual adjustments to achieve good results \cite{mukherjee2022retinal,he2019deep}. For example, segmentation was frequently framed as an optimization problem modeled after graph theory to determine the most likely layer boundaries \cite{garvin2009automated}. These approaches used handcrafted features such as intensity gradients and edge information derived from OCT images.

Despite providing a baseline, these approaches had significant limitations, particularly in patients with severe retinal abnormalities. Conditions such as late AMD or DME can distort retinal layers due to fluid or deposits, complicating boundary detection for traditional algorithms \cite{mukherjee2022retinal}. Additionally, inherent image challenges like speckle noise and low contrast negatively affected the effectiveness and reliability of these methods \cite{mishra2009intra,yazdanpanah2010segmentation}. Though alternative algorithms, such as active contour models and kernel-based optimizations, were developed to handle complex cases, they still required manual input and lacked adaptability, limiting their clinical applicability. These challenges motivated increased interest in data-driven and automated approaches.

\subsection{Deep Learning in Medical Image Segmentation}
Deep Learning (DL) has significantly improved medical image segmentation by providing an automated, efficient, and highly accurate alternative to traditional methods. OCT imaging has gained strong attention for detecting retinal pathology using DL models to segment retinal layers. These models address typical challenges in OCT data, such as speckle noise, low contrast, and complex anatomical distortions \cite{viedma2022deep,pekala2019deep,ngo2019deep}. Encoder-decoder architectures such as U-Net and SegNet, along with their derivatives, allow accurate segmentation of complex retinal structures.

U-Net, introduced by Ronneberger et al. \cite{ronneberger2015unet}, has become a key model for medical image segmentation. Its symmetric encoder-decoder structure with skip connections supports detailed feature extraction and accurate boundary localization. Variants such as DRUNET have achieved robust results in optic nerve head tissue segmentation \cite{devalla2018drunet}. Residual blocks (Deep Residual U-Net), attention mechanisms (Multi-Attention Gated Residual U-Net), and recurrent convolutional layers (Recurrent Residual U-Net) have further enhanced performance in complex tasks \cite{li2019optical,hussain2024magres,alom2019recurrent}.

SegNet, introduced by Badrinarayanan et al. \cite{badrinarayanan2017segnet}, is valued for efficiency and low memory usage, making it suitable for clinical environments requiring rapid processing. Unlike U-Net, SegNet uses encoder pooling indices during decoding, preserving spatial details while saving memory. Studies show SegNet can achieve accuracy comparable to U-Net with fewer parameters \cite{jiang2024analysis,lou2023multiscale}. Recent improvements include deformable convolutions and attention modules to better handle anatomical variation and focus on clinically relevant regions \cite{fang2016segmentation,mani2024moderated}. Hybrid architectures such as CM-SegNet combine convolutional layers with multilayer perceptrons for high-dimensional, multi-modal imaging \cite{xing2022cmsegnet}. Despite these advancements, annotated data limitations remain a hurdle, motivating research into lightweight attention modules and self-supervised learning.

\subsection{Explainable Artificial Intelligence (XAI) in Medical Image Segmentation}
Explainable Artificial Intelligence (XAI) refers to systems whose outputs or processes are transparent and understandable to humans \cite{ali2023explainable}. In medical imaging, where AI predictions directly impact diagnosis and treatment, explainability is crucial \cite{bonazzola2024unsupervised}. Deep learning algorithms, despite their high performance, are considered black-box models, leading to clinician distrust when model predictions contradict clinical judgment \cite{jordan2024estimation}.

XAI presents AI decision-making processes via visual heatmaps, numerical feature importance scores, or rule-based explanations, aiding clinicians in validating segmentation accuracy \cite{ali2023explainable}. In brain MRI, XAI ensures models identify tumor boundaries rather than artifacts \cite{he2019deep}. In OCT imaging, it ensures AI boundaries correspond to anatomical structures \cite{hardani2024decoding}. Gradient-weighted Class Activation Mapping (Grad-CAM) is a widely used XAI technique that produces heatmaps highlighting regions impacting predictions \cite{selvaraju2017gradcam}. Adaptations like Seg-HiRes-Grad CAM enable pixel-wise explanations \cite{rheude2024leveraging}, and architectures such as Double U-Net and R2U-Net incorporate Grad-CAM to provide interpretable visual feedback \cite{xiao2021visualization}.

Challenges to clinical adoption include computational overhead and subjective interpretation of visual explanations. Standardized interpretation guidelines are essential \cite{selvaraju2017gradcam}. In summary, XAI bridges the gap between AI segmentation performance and clinical usability, with techniques like Grad-CAM and attention mechanisms building trust in AI-assisted diagnosis and treatment planning.

\section{Methodology}
\subsection{Dataset Preparation and Preprocessing}
\begin{figure*}[ht]
    \centering
    \includegraphics[width=\textwidth]{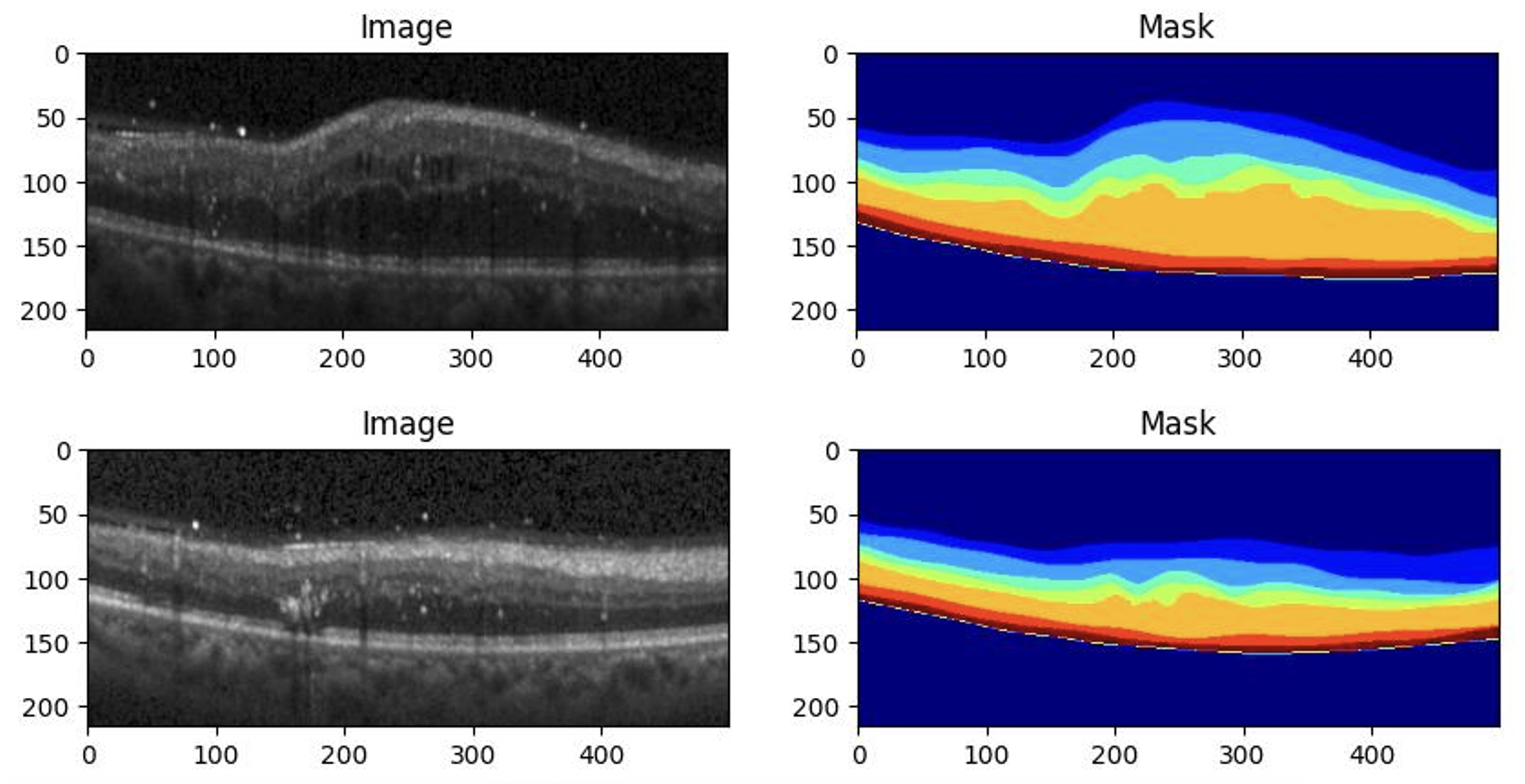}
    \caption{\centering Two random samples (grayscale images) with their respective segmentation masks (color-coded using jet colormap).}
    \label{fig:dataset}
\end{figure*}

The dataset used in this study was comprised of retinal Optical Coherence Tomography (OCT) images and their underlying segmentation masks from the Duke OCT dataset. These images showcase detailed structure from specific retinal layers and are essential for diagnosing several ocular diseases. As shown in Figure~\ref{fig:dataset}, the first step after loading the dataset was to perform exploratory data analysis (EDA) to examine the images and segmentation masks. The dataset consisted of 220 images, having a resolution of 216 × 500 pixels. Images were saved and stored as int64 and the segmentation masks as float64.

Interestingly, a closer look at the mask values showed that there were 8 unique values (0-7) in them, each corresponding to a particular retinal layer.  This confirms that the problem is a multi-class segmentation task, i.e., classifying each pixel to one of the predefined retinal layers. Understanding these characteristics of the dataset was key to building a segmentation model that is effective, as it determines loss function, evaluation metric used and network architecture.

The first step was to normalise the pixel intensity values for the images to the range [0,1]. Normalizing the inputs to the model in this way helped to standardize the input data, and prevented extreme variations in the intensity of the pixels from negatively affecting the training process. Subsequently, both the retinal images and its segmentation masks were resized to a common dimension of 256 × 256 pixels. The resizing step was essential to ensure that all input images had a fixed size which would be processed by the deep learning model while preserving spatial information.
The segmentation masks were converted to a one-hot encoded format in order to allow multi-class segmentation. One-hot encoding transformed each pixel value into a separate channel corresponding to different retinal layers. This allows the model to distinguish between multiple classes rather than performing a simple binary classification. This dataset was then split into a training (80\%) and validation (20\%) subsets to test for model performance. The model’s weights were optimized on the training set, while its ability to generalize was measured on the validation set.

\subsection{Model Architecture}

\begin{figure*}[ht]
    \centering
    \includegraphics[width=\textwidth]{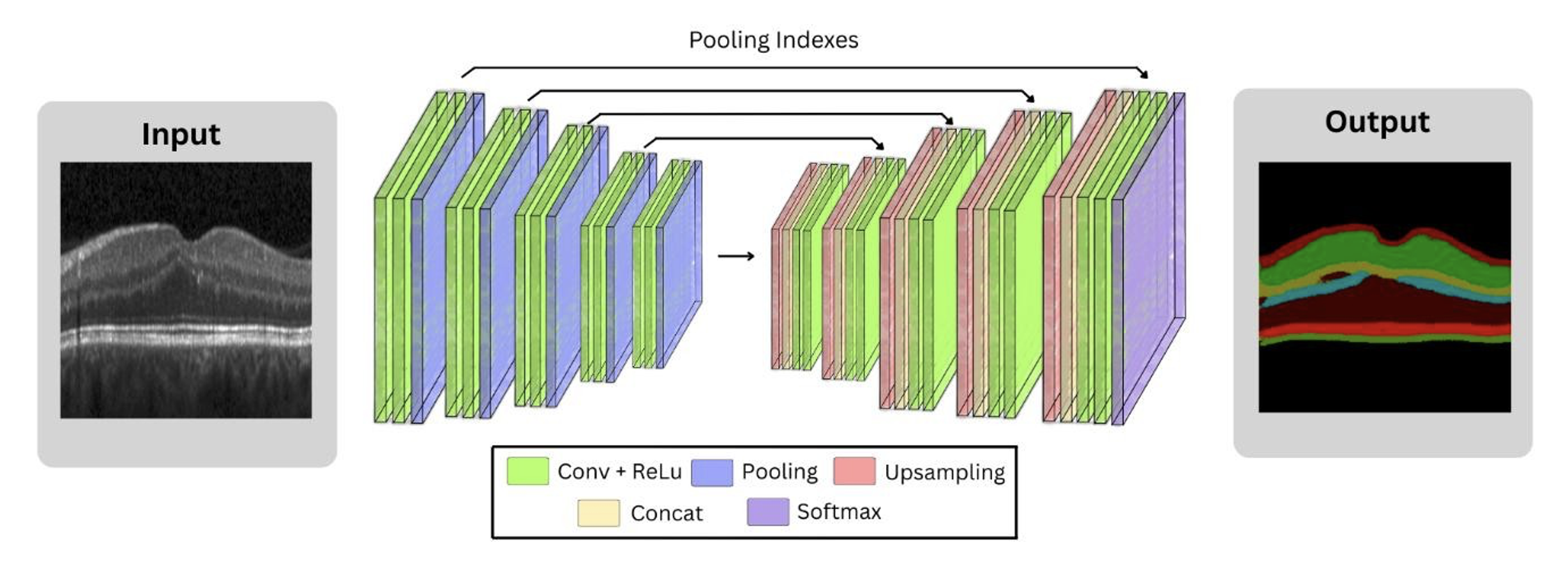}
    \caption{Overview of the SegNet architecture used for retinal layer segmentation, which processes grayscale OCT images through an encoder-decoder module.}
    \label{fig:architecture}
\end{figure*}

This study employs SegNet architecture. SegNet is a specific architecture for pixel-wise image segmentation based on fully convolutional neural network (CNN). Its encoder-decoder architecture allows the model to capture the broad spatial features as well as fine-grained details. Unlike traditional CNNs, SegNet stores pooling indices during downsampling. This enables an accurate reassembly of spatial details during the upsampling of the image which is crucial for identifying thin and closely packed retinal layers. Figure~\ref{fig:architecture} provides an overview of the SegNet model used in this study. The model input is a single channel (grayscale) retinal image with a dimension of 256×256 pixels. This image is fed to the network where several convolutional and pooling layers are used to extract features. Upsampling layers then reconstruct a segmentation mask which labels each pixel into one of eight retinal layers.

\subsubsection{Encoder: Feature Extraction Through Convolution and Pooling}
The encoder is responsible for feature extraction. Each block contains two convolutional layers and one max-pooling layer. In the convolutional layers, we use $3\times3$ kernels with ReLU activations, and `same' padding preserves the spatial dimensions. Each CNN block extracts increasingly abstract features, with the number of filters starting at 64 and increasing to 512. Max-pooling retains key spatial information while lowering the image size. Unlike many CNNs, SegNet stores the locations of these pooled features (the pooling indices). The decoder then uses these to restore spatial accuracy when upsampling. By the time the encoder is finished, the image is compressed from $256\times256$ to $8\times8$ pixels. Despite this reduction, essential features are retained and, due to the stored indices, spatial accuracy is not lost.

\subsubsection{Decoder: Reconstruction with Upsampling}
The decoder uses progressive upsampling of the compressed feature maps to reconstruct the segmentation mask. SegNet’s decoder uses the pooling indices saved during the encoder phase as part of the upsampling, rather than standard interpolation-based upsampling techniques, which helps restore spatial information with higher accuracy. The decoder block consists of several components that work together to refine the upsampled output. The first layer expands the feature maps by a factor of 2 using a transposed convolutional layer (\texttt{Conv2DTranspose}) to reverse the down-sampling carried out by max-pooling. In the next stage, skip connections are applied through concatenation operations. These combine the matching encoder feature maps with the decoder’s layers, helping recover fine details that might have been lost during downsampling. Lastly, two convolutional layers (\texttt{Conv2D}) are applied to further refine the feature maps in order to improve the accuracy of the segmentation output.

\subsubsection{Output Layer: Multi-Class Segmentation}
The final layer of the network is for multi-class segmentation. A $1\times1$ convolutional layer (\texttt{Conv2D}) is applied to reduce the dimensionality of the feature maps to eight channels, each representing one of the segmented retinal layers. This layer is followed by a softmax activation function that assigns a probability score for every pixel in all eight classes. This ensures that each pixel belongs to one and only one segmentation category. Specifically, the \texttt{Conv2D(8, 1$\times$1)} layer maps the pixels to the eight retinal layers, and the softmax activation converts this mapping into normalized probability distributions for each pixel in the image.

\subsection{Loss Function and Evaluation Metrics}
A custom hybrid loss function combining Categorical Cross-Entropy (CCE) and Dice loss was implemented to ensure accurate segmentation of retinal layers. This combination improves pixel-wise classification accuracy and overall segmentation quality. The loss function is a critical component in deep learning model training, allowing the model to minimize errors and learn useful patterns from the data.  

For multi-class classification problems, categorical cross-entropy (CCE) is widely used, as each pixel represents one of many classes. The formula for CCE loss is:

\begin{equation}
L_{CCE} = - \sum_{i=1}^{N} y_i \log(\hat{y}_i)
\end{equation}

In equation (1), $y_i$ is the true class label and $\hat{y}_i$ is the predicted probability for each class. This loss ensures that the predicted probability distribution is consistent with the true labels. However, CCE can struggle with class imbalance, which is common in medical imaging where some retinal layers are smaller than others.  

To address this, Dice loss was added. Dice loss is derived from the Dice coefficient, which measures the overlap between the predicted mask $A$ and ground truth mask $B$. The Dice coefficient is defined in equation (2):

\begin{equation}
\text{Dice Coefficient} = \frac{2 |A \cap B|}{|A| + |B|}
\end{equation}

The hybrid loss function combines CCE and Dice loss as follows in equation (3):

\begin{equation}
L = L_{CCE} + 0.5 \times (1 - \text{Dice Coefficient})
\end{equation}

This allows the model to focus on both pixel-wise classification accuracy and segmentation of overlapping regions.  

\subsubsection{Evaluation Metrics}
Model performance was evaluated using accuracy, Dice coefficient, and Jaccard Index (Intersection over Union, IoU).  

Accuracy is defined as the proportion of correctly classified pixels across all retinal layers, as formulated in equation (4):

\begin{equation}
\text{Accuracy} = \frac{\text{Correctly Classified Pixels}}{\text{Total Pixels}}
\end{equation}

While accuracy is a good general measure, it does not account for class imbalance. The Jaccard Index measures how closely the predicted segmentation matches the true mask and is defined as:

\begin{equation}
\text{IoU} = \frac{|A \cap B|}{|A \cup B|}
\end{equation}

From equation (5), IoU is stricter than the Dice coefficient regarding false positives, making it particularly useful for evaluating segmentation precision.  

The custom hybrid loss function combining Categorical Cross-Entropy and Dice loss enhances pixel-wise classification accuracy and segmentation quality by addressing class imbalance. Using this hybrid function, the model is expected to yield strong results, as evaluated by accuracy, Dice coefficient, and Jaccard Index.

\subsection{Training Procedure}
The model was trained with an initial learning rate of 0.001 using the Adam optimizer. Adam was chosen due to its adaptive learning rate mechanism, which dynamically adjusts the step size for each parameter. This enables faster convergence than traditional optimizers such as Stochastic Gradient Descent (SGD), prevents large oscillations in weight updates, and smooths the gradients. The learning rate of 0.001 was chosen to balance convergence speed and divergence.  

Training was performed for 100 epochs using a batch size of 32, a value that balances training stability and computational efficiency. A batch size of 32 allows gradient updates based on a sufficient number of samples to stabilize learning while avoiding excessive memory usage. Over 100 epochs, the model was able to learn meaningful representations without underfitting. Several callback functions were employed to ensure optimal generalization and avoid overfitting:

\begin{enumerate}
    \item \textbf{ReduceLROnPlateau:} Monitors the validation loss and reduces the learning rate when the loss does not improve for a specified number of epochs, allowing the model to fine-tune parameters with a lower learning rate.
    \item \textbf{EarlyStopping:} Stops training automatically if validation loss does not improve for a specified number of epochs, preventing overfitting by stopping before the model memorizes the training data.
    \item \textbf{ModelCheckpoint:} Saves the model weights with the least validation loss to retain the best performing model, preventing performance degradation in later epochs.
    \item \textbf{CSVLogger:} Saves key training metrics after every epoch, including training loss, validation loss, accuracy, Dice coefficient, and IoU, allowing performance tracking and insights for hyperparameter tuning.
\end{enumerate}

These techniques allowed the model to effectively learn complex patterns in retinal layer segmentation while balancing training speed, generalizability, and overfitting prevention.

\subsection{Model Evaluation and Visualization}
The trained model was evaluated both quantitatively and qualitatively. Quantitative segmentation quality was assessed using accuracy, Dice coefficient, and Jaccard Index on the validation set. Class-wise performance was also examined to determine if specific retinal layers were more challenging to segment accurately.  

Qualitative evaluation involved visually comparing predicted segmentation masks with ground truth masks. This allowed assessment of the model’s ability to retain fine details and accurately delineate different retinal layers. Misclassification analysis highlighted areas where the model deviated from the ground truth, helping to identify patterns in segmentation errors, such as confusion between adjacent layers, which could inform future refinements.

\subsection{Experimental Setup}
The study was conducted on the Kaggle cloud platform using free-tier resources: NVIDIA Tesla P100 GPU (16 GB VRAM), 57.6 GiB RAM, and Python 3.10.12 with pre-installed libraries including TensorFlow 2.12.0 (GPU), Keras, and OpenCV. The Duke OCT dataset of 220 volumes was used, resizing images to $256\times256$ pixels using bilinear interpolation and masks with nearest-neighbor interpolation to preserve pathological features, without augmentation.  

Due to GPU memory limitations, a modified SegNet architecture was used. To improve computational and memory efficiency, the maximum filter depth was set to 512, reflecting a memory-efficient architecture that performs similarly to a 1024-filter version. Full-resolution skip connections were incorporated throughout the network to retain spatial information critical for segmenting thin retinal layers. The network ends with a softmax activation layer for precise multi-class segmentation of every retinal layer.  

Training used a batch size of 32 to fully utilize GPU capacity, and the Adam optimizer with learning rate 0.001, $\beta_1 = 0.9$, and $\beta_2 = 0.999$. To balance pixel-level classification accuracy with segmentation quality, a hybrid loss function was used with equal weighting ($\lambda = 0.5$) of categorical cross-entropy and Dice loss.

\subsection{Explainable AI}
Explainable AI (XAI) integration into deep learning-based Optical Coherence Tomography (OCT) retinal layer segmentation is crucial for enhancing interpretability and clinical reliability. The high segmentation accuracy of deep neural networks is often overshadowed by their black-box behavior, which hinders medical professionals' trust and adoption in routine clinical work. This research develops a SegNet-based segmentation system with integrated Gradient-weighted Class Activation Mapping (Grad-CAM) to ensure that model predictions align with clinically relevant retinal biomarkers. Unlike traditional segmentation models that provide pixel-wise classifications without justification, the proposed approach generates heatmaps for distinct retinal layers to show which areas affect the class distribution. The transparency of this method is essential both for result verification and for detecting structural biases, ensuring the AI system focuses on the correct anatomical structures.

Grad-CAM operates by computing the gradient of the model’s predicted segmentation output with respect to feature maps in a selected convolutional layer, chosen for its high-level spatial representations. The importance of each feature map, $A_k$, for a given class $c$ is determined by the gradient-weighted activation:

\begin{equation}
\alpha^c_k = \frac{1}{Z} \sum_i \sum_j \frac{\partial Y^c}{\partial A^k_{ij}}
\end{equation}

In equation (6), $Y^c$ represents the class-specific output, $Z$ is the spatial dimension of the feature map, and $\frac{\partial Y^c}{\partial A^k_{ij}}$ denotes the gradient of the class score with respect to the activation $A^k_{ij}$. The resulting Grad-CAM heatmap is then computed as:

\begin{equation}
L^c_{\text{Grad-CAM}} = \text{ReLU} \Big( \sum_k \alpha^c_k A_k \Big)
\end{equation}

From equation (7), only positive contributions pass through the ReLU function, which prevents irrelevant activations. This approach is applied to each of the eight segmented retinal layers to generate detailed heatmaps that enhance model interpretability. The mean intensity of the Grad-CAM heatmap provides an overall measure of how strongly the model focuses on a given class across the image, calculated as:

\begin{equation}
\text{Mean Heatmap Intensity} = \frac{1}{N} \sum_{i=1}^{N} L^c_{\text{Grad-CAM}}(i)
\end{equation}

In equation (8), $N$ is the total number of pixels in the heatmap and $L^c_{\text{Grad-CAM}}(i)$ represents the activation value at pixel $i$ for class $c$.  

The Grad-CAM process involves extracting feature maps from a selected layer, calculating gradients of the segmentation output with respect to these maps, performing global average pooling to find channel importance ($\alpha^c_k$), generating the spatial heatmap ($L^c_{\text{Grad-CAM}}$), and overlaying this heatmap on the original OCT scan for interpretation.

A fundamental challenge in medical image segmentation is multi-class explainability, since multiple retinal layers exist within a single OCT scan. Unlike classification models that produce a single class-based heatmap, this research extends Grad-CAM functionalities to all segmented layers, providing specific heatmaps for each anatomical region. Through its multi-class capability, ophthalmologists can monitor the alignment of each layer with expert annotations. When segmentation errors occur, manual validation can guide model refinements, ensuring better consistency across diverse datasets.


\section{Experimental Results}

\subsection{Model Training Performance and Evaluation}
\begin{figure*}[ht]
    \centering
    \includegraphics[width=\textwidth]{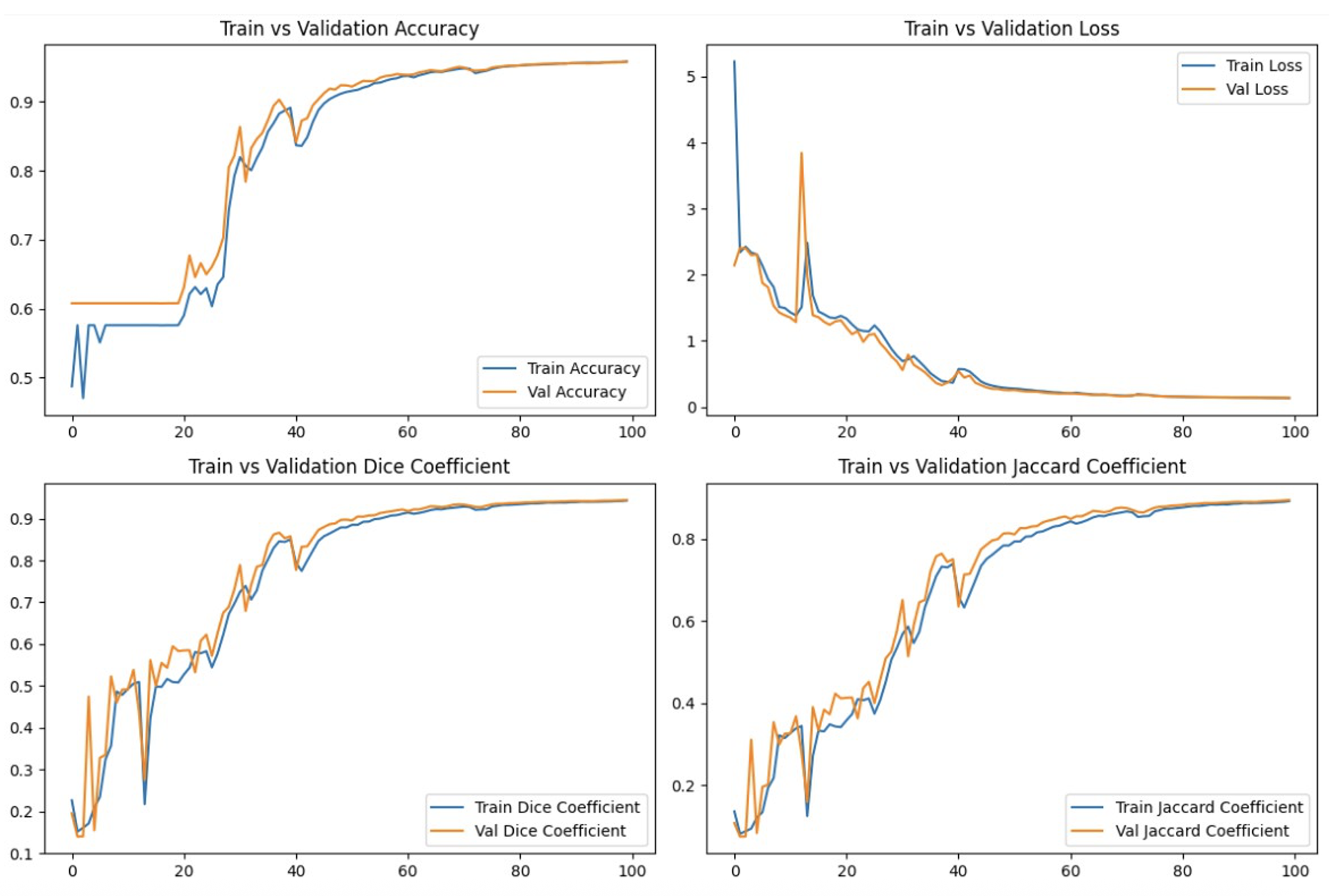}
    \caption{Training and validation performance metrics over epochs.}
    \label{fig:training_history}
\end{figure*}

Extensive training was conducted using OCT images to assess the effectiveness of the proposed model. The training process was monitored using key performance metrics, including Accuracy, Loss, Dice Coefficient, and Jaccard Index (IoU). These metrics are widely recognized in medical image segmentation tasks. Figure~\ref{fig:training_history} illustrates the training history, showing the progression of these metrics across successive epochs for both the training and validation sets.

The training curves provide essential insights into the model's learning behavior, convergence, and generalization ability. The accuracy trend shows consistent improvement, achieving 95.77\% for both training and validation, indicating that the model effectively learns fundamental patterns and generalizes well. This convergence demonstrates robust generalization capacity and reduces the likelihood of overfitting.  

The loss function exhibits a steady decline, with final training and validation losses of 0.1359 and 0.1354, respectively. The comparable values further substantiate the model's robustness and consistent performance on unseen data. The absence of significant divergence between training and validation losses indicates well-balanced learning and effective optimization.  

The Dice Coefficient, crucial for assessing the overlap between predicted and actual segmentations, shows consistent improvement. The final Dice scores of 0.9445 (training) and 0.9446 (validation) illustrate the model's proficiency in precisely delineating fine-grained structures, essential for medical image segmentation tasks. Similarly, the Jaccard Index (IoU) exhibits a final training value of 0.8948 and validation value of 0.8951, reinforcing the reliability of the segmentation. The significant overlap between predicted masks and ground truth confirms the model's ability to produce accurate segmentation results, crucial for clinical decision-making.  

Table~\ref{tab:performance_metrics} provides quantitative results derived from essential segmentation metrics for both training and validation datasets, further validating the model's performance.

\begin{table}[h]
  \centering
  \caption{Quantitative performance of the proposed model on training and validation sets.}
  \label{tab:performance_metrics}
  \begin{tabular}{lcc}
    \toprule
    \textbf{Metric} & \textbf{Training Set} & \textbf{Validation Set} \\
    \midrule
    Accuracy              & 95.77\% & 95.77\% \\
    Dice Coefficient      & 0.9445  & 0.9446  \\
    Jaccard Index (IoU)   & 0.8948  & 0.8951  \\
    Loss                  & 0.1359  & 0.1354  \\
    \bottomrule
  \end{tabular}
\end{table}

\subsection{Visual Interpretation of Prediction Results}

\begin{figure}[h]
    \centering
    \includegraphics[width=\textwidth]{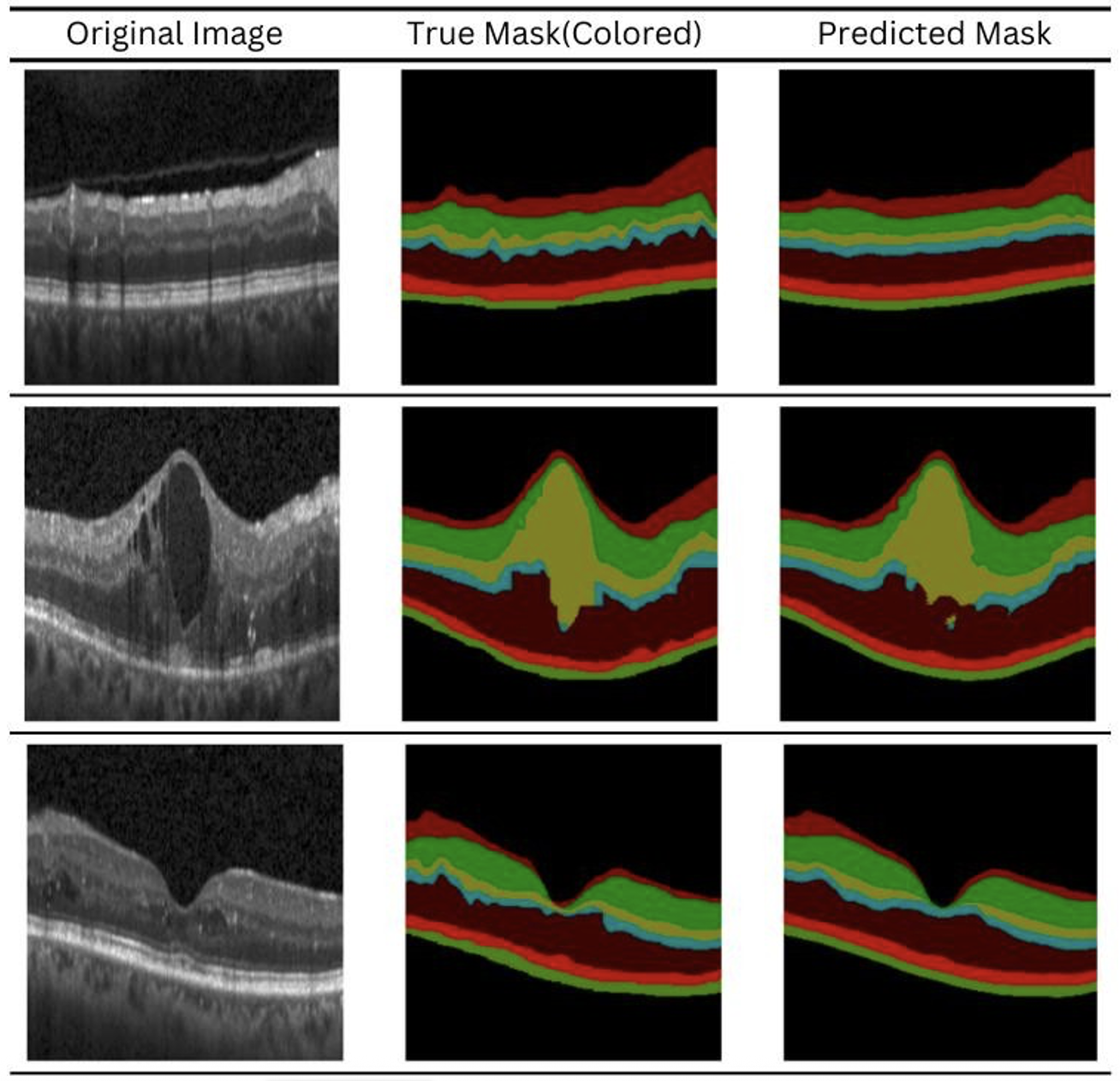}
    \caption{Visual comparison of OCT input, ground truth, and predicted retinal layer segmentation masks across three different OCT samples.}
    \label{fig:visual_results}
\end{figure}

Figure~\ref{fig:visual_results} presents a comparison of the input OCT image, ground truth mask, and the predicted segmentation output, demonstrating the model's ability to accurately differentiate retinal structures. The notable indicators of segmentation performance include spatial alignment, boundary adherence, and structural coherence between the predicted and actual segmentation masks. A detailed visual assessment indicates that the model captures major anatomical regions effectively, exhibiting strong spatial consistency and well-defined boundary segmentation. However, minor deviations are observed in complex areas, particularly in low-contrast regions where adjacent retinal layers have similar intensity distributions, generating potential boundary ambiguities.  

Fine-grained structures, such as thin-layer boundaries and tiny lesions, posed challenges due to the model's receptive field limitations, occasionally leading to minor misrepresentations. Texture-sensitive regions, defined by high-frequency variations, may introduce segmentation uncertainty, false positives, or minor structural inconsistencies. Segmentation performance is strongly influenced by feature representation learning. Regions with distinct intensity gradients and well-defined textures are segmented with high precision, demonstrating the model's effectiveness in capturing strong feature variations. High-contrast edges and structurally homogeneous regions are precisely delineated, indicating effective feature extraction.  

However, segmentation performance may degrade in areas with overlapping textures or inter-class variability, leading to localized inaccuracies. Such errors often arise when the model relies on global contextual information rather than fine-grained spatial details, occasionally resulting in over-segmentation of adjacent regions or under-segmentation in subtle boundary areas. A detailed analysis of misclassified pixels reveals three primary failure modes: border ambiguity, over-segmentation, and under-segmentation. Boundary uncertainty occurs in fine-grained structures where loss functions prioritize global structure over precise edges, producing slightly blurred contours. Integration of boundary-aware losses (e.g., Active Contour Loss) could enhance edge fidelity. Incorporating hybrid attention mechanisms alongside boundary-aware losses could further mitigate these errors by enhancing the model’s focus on subtle structural cues while preserving global contextual consistency.

High-variance regions generate over-segmentation, where excessive feature activation produces false positives. Under-segmentation affects rare-class areas, where data imbalance causes the model to struggle with small-scale structures. Class-weighted loss functions (e.g., Focal Loss) or data augmentation could improve segmentation accuracy in these cases. Although the model excels in capturing high-level spatial patterns, future developments in multi-scale feature fusion, uncertainty estimation, and self-supervised pre-training could enhance fine-grained feature encoding, ensuring more accurate segmentation in ambiguous and underrepresented areas.  

To further validate these qualitative observations, robust quantitative metrics such as Intersection over Union (IoU) and Dice Coefficient were used to provide a comprehensive assessment of segmentation accuracy. This combined analysis ensures that the model not only performs well visually but also achieves statistically reliable segmentation, supporting its potential use in real-world clinical contexts.

\subsection{Class-wise IoU Scores and Segmentation Accuracy}

To assess the performance of the segmentation model, both segmentation accuracy and Intersection over Union (IoU) metrics were evaluated across different segmentation classes. These metrics provide valuable insights into the model’s ability to correctly classify and delineate segmented regions. Table~\ref{tab:classwise_iou} presents the class-wise IoU scores and segmentation accuracy, offering a detailed view of how well the model performs across different segmentation classes.

\begin{table*}[hbt!]
  \centering
  \caption{Class-wise IoU scores and segmentation accuracy for the proposed model.}
  \label{tab:classwise_iou}
  \begin{tabular}{lcccccccc}
    \toprule
    \textbf{Segmentation Class} & \textbf{0} & \textbf{1} & \textbf{2} & \textbf{3} & \textbf{4} & \textbf{5} & \textbf{6} & \textbf{7} \\
    \midrule
    IoU Score                  & 0.99 & 0.82 & 0.86 & 0.71 & 0.68 & 0.90 & 0.82 & 0.78 \\
    Segmentation Accuracy (\%)  & 99.2 & 91.3 & 93.7 & 85.2 & 80.5 & 95.1 & 91.7 & 88.6 \\
    \bottomrule
  \end{tabular}
\end{table*}

The model demonstrates exceptional segmentation accuracy and IoU scores for Class 0, attaining an IoU of 0.99 and a segmentation accuracy of 99.2\%, signifying nearly flawless demarcation. Classes 2, 5, and 6 exhibit robust performance, achieving IoU scores surpassing 0.85 and accuracy exceeding 90\%, validating the model's capability of accurately segmenting these retinal layers with precision. Classes 1 and 7 exhibit moderate performance, with IoU values between 0.78 and 0.82, indicating reliable segmentation that could benefit from further enhancement. Classes 3 and 4 exhibit diminished IoU scores of 0.71 and 0.68, along with decreased segmentation accuracy, highlighting challenges in differentiating these layers.  

\begin{figure*}[!hbt]
  \centering
  \includegraphics[width=\textwidth]{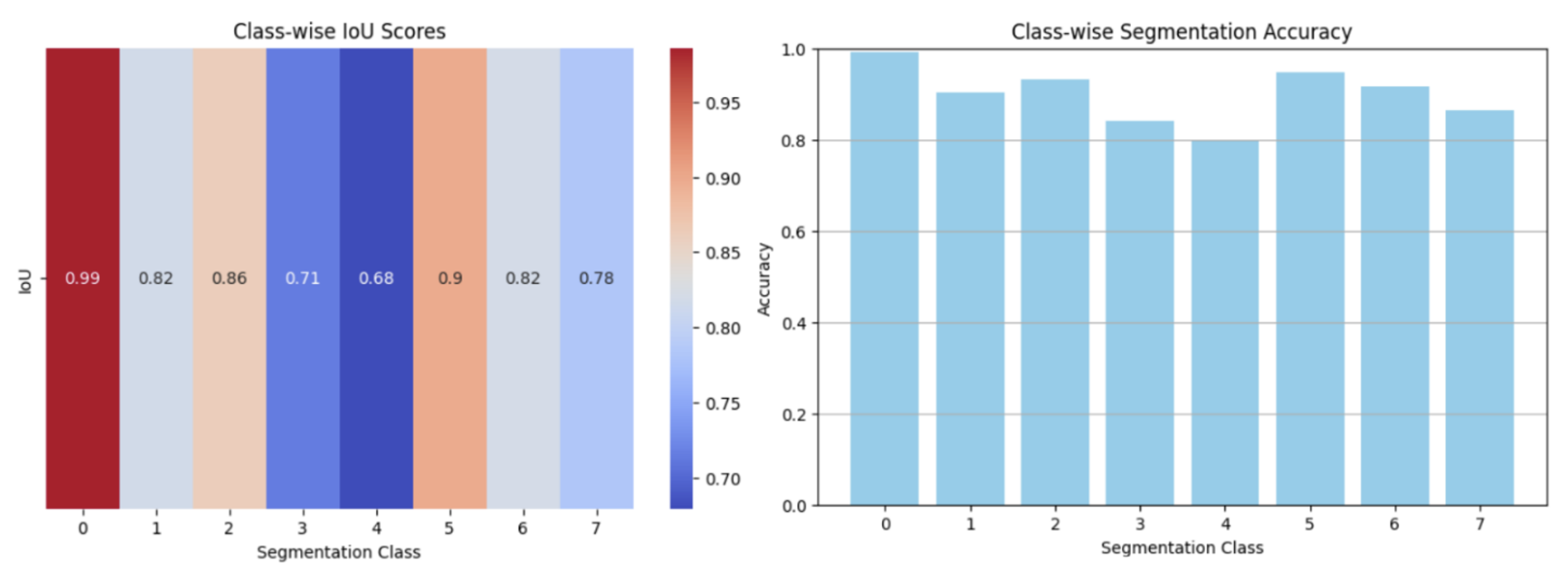}
  \caption{Left: Class-wise IoU scores for segmentation performance. Right: Segmentation accuracy across different classes.}
  \label{fig:iou_accuracy}
\end{figure*}

The lower performance in some classes could be attributed to class imbalance, complex anatomical structure, and overlapping boundaries. Integration of data augmentation, attention mechanisms, and specialized loss functions (e.g., Dice loss or Focal loss) could enhance model robustness, especially in challenging regions. Figure~\ref{fig:iou_accuracy} (left) illustrates a heatmap displaying the class-wise IoU scores for segmentation performance, where the color gradient highlights variations in IoU scores with elevated values depicted in red and lower values in blue. Figure~\ref{fig:iou_accuracy} (right) shows the segmentation accuracy across different classes. These visualizations offer a clear understanding of the model's performance variances among classes. Despite minor inconsistencies, the overall high IoU and segmentation accuracy validate the model’s reliability in multi-class segmentation, reinforcing its applicability in real-world scenarios.

\subsection{Explainable AI (XAI) Analysis for Retinal Layer Segmentation}

The analysis of explainability for the proposed SegNet-based retinal layer segmentation model focuses on understanding hierarchical feature learning via two principal convolutional layers: \texttt{Conv2d\_19} and \texttt{Conv2d\_20}. \texttt{Conv2d\_19}, a feature refinement layer within the decoder pathway, assimilates fine-grained spatial data ensuring accurate segmentation boundaries. Grad-CAM visualizations illustrate its role in identifying edge structures and textural patterns crucial for differentiating retinal layers.  Conversely, \texttt{Conv2d\_20}, the ultimate segmentation layer, produces an 8-class pixel-wise classification, directly correlating learned features to specific retinal structures. Its activations facilitate class-specific decision mapping, highlighting semantic alignment while maintaining confidence and spatial precision. The comparative study of these layers underscores the model's interpretability and clinical relevance, ensuring dependable AI-driven retinal segmentation. 

A primary challenge in explainable artificial intelligence for segmentation models is handling multiple classes within a singular output. Unlike classification models, our approach extends Grad-CAM to generate class-specific heatmaps, enabling detailed per-layer verification, failure mode analysis, and cross-scan generalization. This guarantees precise segmentation of inner and outer retinal layers while detecting potential class misalignments.  The comparative analysis of the heatmaps for \texttt{Conv2d\_19} and \texttt{Conv2d\_20} reveals distinctions in feature refinement, class specificity, and interpretability, providing significant insights into the model's decision-making process and clinical significance. These findings highlight how explainability not only enhances model transparency but also fosters greater trust and adoption in clinical settings.

\begin{figure*}[hbt!]
  \centering
  \includegraphics[width=\textwidth]{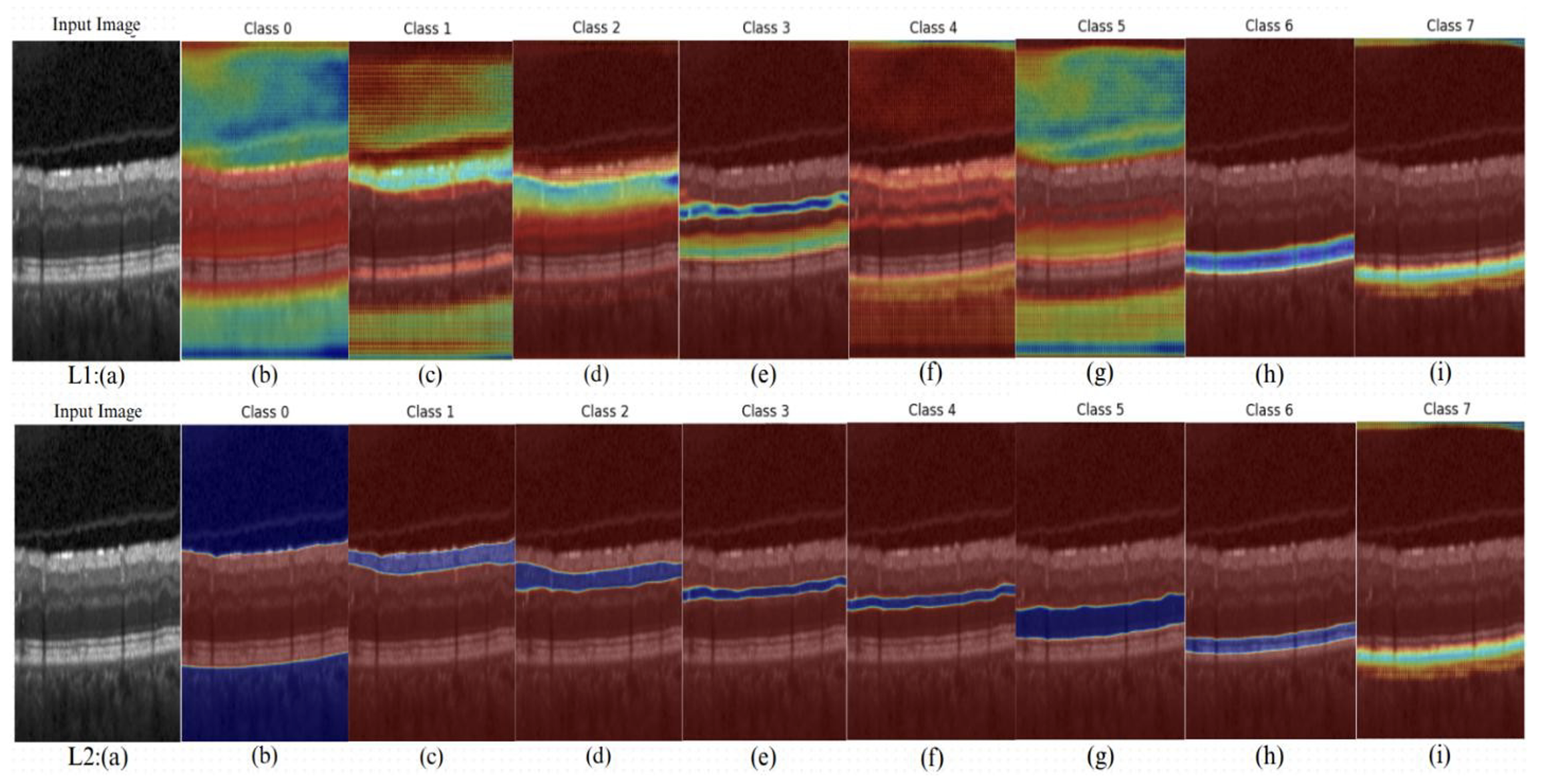}
  \caption{Visualization of hierarchical feature maps from two consecutive convolutional layers (Conv2d\_19 and Conv2d\_20). (L1): (a) Input image; (b--i) Feature maps from Conv2d\_19, corresponding to Class 0--7. Bottom row (L2): (a--i) Feature maps from Conv2d\_20.}
  \label{fig:gradcam_heatmaps}
\end{figure*}

The Grad-CAM visualizations presented in Figure~\ref{fig:gradcam_heatmaps} illustrate the activation patterns of the model across different retinal structures, offering insights into its feature learning process for class predictions. In the top row (L1), the heatmaps (b–i) reveal varying attention distributions for the convolutional layer \texttt{Conv2d\_19}, with certain classes (Class 2, Class 3, and Class 6) demonstrating localized activations along specific retinal layers, suggesting precise feature recognition. Conversely, other classes, such as Class 0 and Class 5, exhibit more dispersed activations, implying reliance on broader structural features.  

In the bottom row (L2), for the convolutional layer \texttt{Conv2d\_20}, activations appear more refined, with Class 2, Class 3, and Class 6 showing concentrated attention on thin, well-defined regions, reinforcing the model’s ability to detect subtle patterns. Notably, Class 0 exhibits reduced activation in the upper region, as indicated by prominent blue regions, suggesting lower feature importance. The comparative analysis between L1 and L2 indicates that deeper layers (L2) contribute to more precise feature extraction by filtering out irrelevant activations, thereby improving interpretability and enhancing classification accuracy, particularly for structurally distinct retinal abnormalities.

\begin{table*}[hbt!]
  \centering
  \caption{Feature importance, max heatmap activation, and mean heatmap intensity for Conv2d\_19 and Conv2d\_20.}
  \label{tab:feature_weights}
  \resizebox{0.95\textwidth}{!}{%
  \begin{tabular}{l l cccccccc}
    \toprule
    \textbf{Layer} & \textbf{Metric / Class} & \textbf{0} & \textbf{1} & \textbf{2} & \textbf{3} & \textbf{4} & \textbf{5} & \textbf{6} & \textbf{7} \\
    \midrule
    \multirow{3}{*}{\rotatebox[origin=c]{0}{Conv2d\_19}} 
      & Feature Importance Weights ($\alpha_{kc}$) & 9.81e-05 & -2.06e-04 & 1.93e-05 & -7.21e-05 & 9.62e-05 & -6.52e-05 & -4.81e-05 & 4.26e-06 \\
      & Max Heatmap Activation ($L_{\text{Grad-CAM}_c}$) & 1.000 & 1.000 & 1.000 & 1.000 & 1.000 & 1.000 & 1.000 & 1.000 \\
      & Mean Heatmap Intensity & 0.369 & 0.208 & 0.084 & 0.059 & 0.090 & 0.288 & 0.051 & 0.045 \\
    \midrule
    \multirow{3}{*}{\rotatebox[origin=c]{0}{Conv2d\_20}} 
      & Feature Importance Weights ($\alpha_{kc}$) & 0.125 & 0.125 & 0.125 & 0.125 & 0.125 & 0.125 & 0.125 & 0.125 \\
      & Max Heatmap Activation ($L_{\text{Grad-CAM}_c}$) & 1.000 & 1.000 & 1.000 & 1.000 & 1.000 & 1.000 & 1.000 & 1.000 \\
      & Mean Heatmap Intensity & 0.641 & 0.067 & 0.060 & 0.029 & 0.033 & 0.095 & 0.044 & 0.031 \\
    \bottomrule
  \end{tabular}}
\end{table*}

The analysis in Table~\ref{tab:feature_weights} provides key insights into hierarchical feature extraction through calculated feature importance weights, max heatmap activations, and mean heatmap intensities. \texttt{Conv2d\_19} operates as a feature refinement layer because its weight values vary based on retinal classes to extract vital spatial and structural details needed for accurate segmentation. The variation in mean heatmap intensities suggests that this layer preserves fine-grained information, specifically in complex boundaries such as Class 0 and Class 5, which exhibit higher intensity values.  

The final segmentation layer, \texttt{Conv2d\_20}, exhibits more uniform feature importance values across all classes and is responsible for final class-specific segmentation, indicating its role in refining and aggregating learned features into distinct segmentation outputs. The normalization process generates maximum heatmap activations at 1.000 across both layers, keeping all activations between (0,1). The mean heatmap intensity outcomes from \texttt{Conv2d\_20} demonstrate better class separability, indicating the layer’s ability to enhance prediction reliability.  

Grad-CAM aids in detecting model biases and misclassifications, specifically in cases where overlapping activations indicate confusion between adjacent retinal layers or where incorrect segmentation confidence suggests dataset imbalances. Inconsistencies across scans further highlight the need for robust augmentation techniques to mitigate sensitivity to imaging artifacts. Ultimately, the Grad-CAM analysis of \texttt{Conv2d\_19} and \texttt{Conv2d\_20} confirms the model’s effectiveness in hierarchical feature extraction, refining spatial precision while ensuring accurate pixel-wise classification. The visualization of per-class heatmaps increases transparency, allowing clinicians to validate the AI-based segmentation process used for automatic retinal disease tracking. These findings demonstrate how XAI can be effectively integrated into deep learning frameworks, delivering reliable models alongside clinical transparency and usability in medical imaging.

\section{Conclusion}

Retinal layer segmentation in Optical Coherence Tomography (OCT) plays a vital role in diagnosing and monitoring vision-threatening diseases. This study presents an enhanced SegNet-based deep learning framework that bridges the gap between clinical interpretability and algorithmic accuracy. Key contributions include architectural improvements, a hybrid loss formulation, and the integration of Explainable AI (XAI) methods.  

The proposed model enables precise delineation of fine retinal structures through hybrid pooling indices and adaptive receptive fields, ensuring robust feature extraction even from noisy OCT scans. For thin and overlapping layers, the combination of categorical cross-entropy and Dice loss efficiently mitigates class imbalance, thereby improving segmentation accuracy. By generating per-class Grad-CAM heatmaps, the model further enhances transparency by aligning predictions with clinically recognized anatomical structures, providing clinicians with a visual justification for the segmentation results.  

Empirical testing on the Duke OCT dataset demonstrates strong performance, achieving 95.77\% validation accuracy, a Dice coefficient of 0.9446, and a Jaccard Index (IoU) of 0.8951. These results indicate the model’s ability to generalize across different scans while maintaining computational efficiency, reducing analysis time from minutes to seconds—an important advantage in clinical practice. Class-wise analysis highlights challenges in segmenting Classes 3 and 4 (IoUs of 0.71 and 0.68), attributed to boundary ambiguity and structural complexity. Addressing these limitations using class-weighted loss functions or attention-based refinement could further improve segmentation accuracy in uncertain regions.  

Grad-CAM-based XAI analysis verifies hierarchical feature learning, where earlier convolutional layers preserve spatial details and deeper layers refine semantic representations. Layer-wise interpretability not only confirms anatomical integrity of the segmentations but also increases clinician confidence—a critical requirement for AI adoption in healthcare.  

By combining clear model outputs with strong performance metrics, the proposed solution ensures both technical rigor and clinical relevance. The modular design, cloud-based training, and standardized preprocessing enhance reproducibility and facilitate deployment. Future work should incorporate multi-center datasets to improve generalizability, while transformer-based encoders or uncertainty-aware learning may further strengthen performance in pathological cases. Overall, this study advances automated retinal layer segmentation by integrating deep learning accuracy with clinical interpretability, offering a transparent, scalable solution for AI-assisted retinal diagnostics.

\bibliographystyle{IEEEtran} 
\bibliography{references}    

\end{document}